\documentclass[twocolumn,
pra,showpacs,floatfix,superscriptaddress]{revtex4-2}
\usepackage[latin9]{inputenc}
\usepackage{color}
\usepackage{amsfonts}
\usepackage{amsmath}
\usepackage{amssymb}
\usepackage{graphicx}
\usepackage{esint}

\makeatletter
\@ifundefined{textcolor}{}
{%
 \definecolor{BLACK}{gray}{0}
 \definecolor{WHITE}{gray}{1}
 \definecolor{RED}{rgb}{1,0,0}
 \definecolor{GREEN}{rgb}{0,1,0}
 \definecolor{BLUE}{rgb}{0,0,1}
 \definecolor{CYAN}{cmyk}{1,0,0,0}
 \definecolor{MAGENTA}{cmyk}{0,1,0,0}
 \definecolor{YELLOW}{cmyk}{0,0,1,0}
}

\usepackage{color}
\usepackage{graphics}
\usepackage{epsfig}
\newcommand{\be}{\begin{equation}}
\newcommand{\ee}{\end{equation}}
\newcommand{\bea}{\begin{eqnarray}}
\newcommand{\eea}{\end{eqnarray}}
\newcommand{\bes}{\begin{subequations}}
\newcommand{\ees}{\end{subequations}}
\newcommand{\PT}{\mathcal{PT}}

\newcommand{\tphi}{\tilde{\phi}}

\newcommand{\hw}{\hat{w}}

\newcommand{\RE}{\mathrm{Re}\,}
\newcommand{\IM}{\mathrm{Im}\,}

\newcommand{\rect}{\mathrm{rect}}

\newcommand{\s}{0}
\newcommand{\tl}{\tilde{\lambda}}

\newcommand{\rev}[1]{\textcolor{black}{#1}} 
\newcommand{\rerev}[1]{\textcolor{black}{#1}}

\makeatother

\begin{document}

\title{Perfectly absorbed and emitted currents by complex potentials in nonlinear media}

\author{Dmitry A. Zezyulin}
\affiliation{ 
ITMO University, St. Petersburg 197101, Russia}

\author{
 Vladimir V. Konotop 
}
\affiliation{
	Departamento de F\'isica and Centro de F\'isica Te\'orica e Computacional, Faculdade de Ci\^encias, Universidade de Lisboa, Campo Grande 2, Edif\'icio C8, Lisboa 1749-016, Portugal 
}

\date{\today}
\begin{abstract}
	
Recently it was demonstrated that the concept of a spectral singularity (SS) can be generalized to  waves propagating in  nonlinear media\rev{, like matter waves or electromagnetic waves in Kerr media}. The corresponding solutions represent nonlinear currents \rev{sustained by a localized linear complex potential in a nonlinear Schr\"odinger equation}. A key feature allowing the nonlinear generalization of a SS is a possibility to reduce a nonlinear current to the linear limit, where a SS has the unambiguous definition. In the meantime,  known examples of nonlinear modes bifurcating from linear spectral singularities are few and belong to the specific class of constant-amplitude waves. Here  we propose to extend the class of nonlinear SSs by incorporating  solutions whose amplitudes are inhomogeneous. We show that the continuation from the linear limit requires a deformation of the complex potential, and this deformation is not unique. Examples include   the deformation preserving the gain-and-loss distribution  and  the deformation   preserving   geometry  of the potential. For the case example of a rectangular potential, we demonstrate that the nonlinear currents can be divided into two types: solutions of the first type bifurcate from the linear spectral singularities, and solutions of the second type cannot be reduced to the linear limit.

\end{abstract}


\maketitle

\section{Introduction}

Spatially localized complex potentials \rev{in the linear Schr\"odinger equation}, subject to specific constraints on their characteristics, can result in total absorption~\cite{CPA1,CPA5,CPA4,CPA2,CPA3} of the incoming  waves or emission of outgoing waves. Physically, such regimes are referred to as coherent perfect absorption (CPA)~\cite{Stone,CPA4,review} or lasing, respectively. Mathematically, the respective  phenomena are related~\cite{Vainberg,Most09,Stone} to spectral singularities (SSs) emerging in the continuous spectra of the underlying complex potentials~\cite{Guseinov}. The one-to-one correspondence between a solution corresponding to a spectral singularity  (i.e., SS-solution) and a wave either incoming to or outgoing from the localized complex potential has been proven in~\cite{Vainberg} (see also \cite{KLV2019}).  \rev{Using this equivalence, in the one-dimensional geometry (which is assumed below), a linear SS-solution can be defined as a nontrivial solution $\phi_0(x)$ of a \rev{stationary Schr\"odinger equation} with  the asymptotic behavior  $\phi_0(x)\to \rho_\pm e^{i \varphi_\pm \pm i k_0 x}$ at  $x\to \pm \infty$,  where  $\rho_\pm$ and $\varphi_\pm$ are real constants, and a nonzero real $k_0$ characterizes the location of the SS in the continuous spectrum. Positive and negative values of $k_0$ correspond to outgoing (laser) and incoming (CPA) wave boundary conditions, respectively.  Therefore any SS corresponds to a peculiar scattering state in the continuous spectrum. In this respect, SSs  are  distinctively different from exceptional points (EPs) \cite{Kato,Heiss}, i.e., non-Hermitian degeneracies emerging from a coalescence of several square-integrable eigenfunctions from the discrete spectrum. }

\rev{Considering electromagnetic waves in nonlinear Kerr media or superfluid flows in Bose--Einstein condensates (BECs), where two-body interactions define the properties of the background state, one has to take into account the nonlinearity of the respective waveguiding  medium. The impact   the nonlinearity  is   described by the  nonlinear Schr\"odinger (NLS) equation (or the Gross-Pitaevskii equation in the case of BEC theory). Then} a question about  the extension of the concepts of CPA and laser to those nonlinear environments   naturally arises. One can think of different implementations of the mentioned generalization. First, one can consider a nonlinear active layer that absorbs or emits waves, which is  however  embedded in a linear medium. In this case \rev{the effect of the nonlinearity will be confined to the localized spatial domain, but} the  emitted or absorbed waves remain  linear, i.e., they  still propagate in the linear medium~\cite{Zharov_nonlin,nonlin-CPA1,nonlin-CPA2,Most_rect17,nonlin-CPA4}.  Alternatively, one can consider a linear active potential operating in a nonlinear medium. In this case,   the incoming or outgoing waves become  nonlinear currents. Conceptually, such a possibility  was elaborated in \cite{ZezKon2016,Mullers2018,ZOK} and validated in experiments with atomic BECs  \cite{Mullers2018} subjected to a localized dissipation.  While there are obvious similarities between linear and nonlinear  perfect absorption or lasing, one can point out four essential distinctions. First, while a linear monochromatic solution corresponding to a SS is characterized only by its frequency (wavelength or wavenumber) and   by the relation between the right and left propagating waves at opposite infinities, in the nonlinear case the field amplitudes  at the infinities must be equal, \rev{because now the frequency depends on the background amplitude (constancy of the frequency is only possible if the background amplitudes are equal at both infinities).} Moreover, in the nonlinear system rescaling of the amplitude can have nontrivial impact on the absorption or lasing.    Second, nonlinear currents can suffer from  modulational instabilities (resulting from the   nonlinearity  of the medium  rather from  the presence of the gain or loss).  Third, nonlinear currents cannot be explained by (only) the interference effects, which are known to be responsible for the coherent perfect absorption of linear waves  \cite{Stone}. Finally, the very concept of a SS is purely linear and cannot be uniquely defined in the nonlinear case, allowing multistable regimes~\cite{multistab}. Therefore,  a key issue in the continuation of the linear perfectly absorbed or emitted currents  to the nonlinear regime    is the possibility to reduce the nonlinear system to a properly defined linear  limit. Since the effective strength of nonlinearity is determined by the amplitude of the field,  the linear limit can be associated with a regime where the field amplitude vanishes. Using this idea,  several examples of CPAs and lasers for nonlinear waves have been elaborated. However, the known examples  are still   limited to few exceptional  cases, including Dirac delta-potentials   and Wadati potentials~\cite{ZezKon2016,ZezKon20,ZezKonSymm20}, where the linear solution is constant-amplitude and its generalization to the nonlinear case is obvious. In a more general situation, where the linear solution is not constant-amplitude, the  nonlinear  generalization   is not that straightforward.  The reason is that both nonlinear currents and linear SS-solutions require a delicate balance between the parameters  of the  potential and of the current itself. Any change    of the wave parameters (e.g., the wavenumber) or rescaling of the amplitude   (which becomes relevant in the nonlinear case) can violate this balance. As a result,  for the persistence of nonlinear currents one      should simultaneously deform the complex potential itself. Therefore, the transition between the linear and nonlinear regimes implies not only the   increase of the field amplitude, but also  a continuous deformation of the potential.  A more detailed study of possible continuations of  linear spectral singularities to the nonlinear medium is the first goal of the present article. We show that the deformation of the potential that enables the continuation to the nonlinear regime is not unique. Two examples that we  will address  in more detail   correspond to the continuation   that leaves the imaginary part of the potential intact and to the continuation that preserves the shape of the potential. In the latter case, the potential has rectangular shape either in linear and nonlinear cases, but the complex  \rev{coefficient that governs the ``strength''   of the  of the rectangular slab}  depends on the background  field amplitude.   

A distinctive feature of  linear and nonlinear perfectly absorbed or emitted currents consists in specific boundary conditions which imply that   sufficiently far from the potential   the solution transforms into a purely incoming (for CPA) or purely outgoing (for laser) wave. In the meantime, some  nonlinear perfectly absorbed currents that satisfy the boundary conditions of this type have already been considered in the literature, say in   \cite{SK,ASK,ZKBO}, where these solutions have been interpreted in the context of the macroscopic  Zeno effect, due to its resemblance to the well-known quantum Zeno effect~\cite{Zeno1,Zeno2,Zeno3}. However, the eventual relation of these solutions to linear and nonlinear spectral singularities  has not been discussed in any detail. A clarification of this issue is the second goal of the present article. Rather remarkably, we find that there exist nonlinear perfectly absorbed currents that cannot be reduced to the linear limit by changing the \rev{ coefficient} of the rectangular potential.  Therefore  perfectly absorbed or emitted nonlinear currents can be classified in two types:  solutions   of the  first type bifurcate from linear spectral singularities, and  the nonlinear currents of the second type do not admit the described linear limit.

The plan of this article is as follows.  The model and main conventions are introduced in Sec.~\ref{sec:model}.   In Sec.~\ref{sec:Wadati} we describe  a continuation of a linear SS-solution to the nonlinear domain, considering the solution amplitude as a bifurcation parameter.  The procedure developed in this section does not affect the gain-and-loss distribution of the complex potential.   In Sec.~\ref{sec:rectang} we propose a different approach to the nonlinear continuation which is illustrated for the case example of the rectangular potential. Section~\ref{sec:nonlin-currents} compares the nonlinear currents bifurcating from the linear limit and the essentially nonlinear currents that cannot be reduced to the linear limit.  The results are summarized in the concluding Sec.~\ref{sec:concl}.  
 
\section{The model} 
\label{sec:model}
We consider a dimensionless field $\Phi(t,x)$ governed by the one-dimensional defocusing NLS equation 
\begin{equation}
	\label{NLS}
i\Phi_t = -\Phi_{xx} + U(x)\Phi + |\Phi|^2\Phi,
\end{equation}
where $t$ and $x$ correspond to the evolution and transverse coordinates, respectively, and  
$U(x)$ is a localized complex potential:
\begin{equation}
	\label{U}
\lim_{x\to \pm \infty } U(x)=0,
\end{equation}
that decays fast enough. \rev {While hereafter we use dimensionless units, it is known that in the BEC theory $\Phi(t,x)$ is a macroscopic wavefunction \cite{Pita} and $t$ is time,  while in the case of a beam propagating in a Kerr medium $\Phi(t,x)$ is a complex amplitude of the transverse electric field and $t$ is a coordinate along the propagation axis \cite{Kivshar}. In  the BEC context,  the complex potential $U(x)$ can be created either by elimination and loading BEC atoms, while for the optical  beam the  potential   is determined by the complex dielectric permittivity of the cubic medium.}

\begin{figure}
	\begin{centering}%
		\includegraphics[width=\columnwidth]{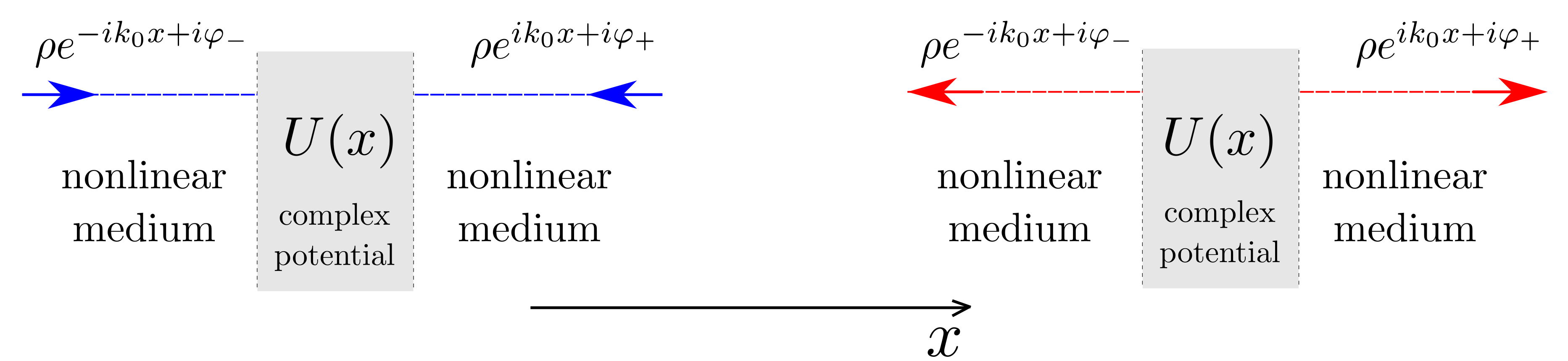}		\par
	\end{centering}
	\caption{\rev{Schematic illustration boundary conditions for the CPA at $k_0<0$ (left panel) and laser at $k_0>0$ (right panel).}  
	}
	\label{fig:0}
\end{figure}

We consider   {\em stationary} solutions $\Phi(t,x) = e^{-i\mu t} \phi(x)$, where $\mu$ is a real parameter \rev{(depending on the physical context, it is a dimensionless chemical potential of the condensate or describes the change of the propagation constant of the optical beam)}. The stationary wavefunction  $\phi$ satisfies  
\begin{equation}
	\label{eq:nonlin}
	-\phi_{xx} + U(x)\phi +  |\phi|^2\phi  = 
	\mu\phi.
\end{equation}
We are interested in solutions of (\ref{eq:nonlin}) subject to the finite density boundary conditions, i.e., when $|\phi(x)|$ tends to a nonzero constant at the infinity. In view of (\ref{U}), the only boundary condition  of this type consistent with (\ref{eq:nonlin}) is $|\phi(x)| \to\rho>0$ at $x\to\pm\infty$. Respectively, we say that $\phi(x)$ is a perfectly absorbed (emitted) current with background  amplitude $\rho$ if there exist real phases    $\varphi_\pm$    and real constant $k_0 <0$ ($k_0 >0$) such that \rev{(see schematics in Fig.~\ref{fig:0})}
\begin{align}
\label{eq:PA}
\lim_{x\to - \infty } \phi(x)e^{ i k_0 x} = \rho e^{i\varphi_-} , \quad \lim_{x\to \infty } \phi(x) e^{- i k_0 x} = \rho e^{i\varphi_+}. 
\end{align}
For such solutions $\mu=\rho^2+k_0^2$.

Generally, a solution of (\ref{NLS}), (\ref{eq:PA}) with fixed amplitude $\rho$ and phases $\varphi_\pm$ 
may exist only for specific choices of the complex potential $U(x)$ which depends on the background amplitude $\rho$. Therefore below we denote such solutions and  potentials by $\phi(x) = \phi_\rho(x)$ and  $U(x)=U_\rho(x)$, respectively,  where subscript $\rho$ emphasizes the dependence   on background amplitude.   Suppose now that there exist the limits 
\begin{align}
	\lim_{\rho\to0}\frac{\phi_\rho(x)}{\rho}=:\phi_\s(x), \quad \lim_{\rho\to 0} U_\rho(x) =: U_0(x).
\end{align} 
Then $\phi_\s(x)$ is a SS-solution of the linear problem
\begin{align}
	\label{eq:linear}
	[-\partial_x^2  + U_0(x)]\phi_\s = k_0^2\phi_\s,
	\\[2mm]
	\label{bound-lin-eq}
\lim_{x\to \pm \infty } \phi_\s(x) e^{ \mp i k_0 x} = e^{i\varphi_\pm}
\end{align}
which will be considered as the definition of the linear limit    of the problem (\ref{NLS}), (\ref{eq:PA}).  
 
Considering the linear limit alone, i.e., not related to the nonlinear problem  (\ref{NLS}), (\ref{eq:PA}), the boundary conditions for (\ref{bound-lin-eq}) can be generalized to unequal amplitudes $\rho_\pm>0$ at the infinities 
\begin{align}
 	\label{bound-lin-uneq}
\lim_{x\to \pm \infty } \phi_\s(x) e^{ \mp i k_0 x} = \rho_\pm e^{i\varphi_\pm}.
\end{align}
Since in the linear problem the field amplitude can be scaled out, although the constants $\rho_+$ and $\rho_-$ are allowed to be different, only the relation between them, $\rho_+/\rho_-$, is a relevant parameter. Thus, given a linear SS solution $\phi_\s(x)$, it can be a limit of the nonlinear absorbing or lasing current only if $\rho_+/\rho_-=1$. Notice, however, that we are considering linear and nonlinear currents corresponding to the same   $k_0$ (i.e., to the same wavelength).

From (\ref{eq:nonlin}) and (\ref{eq:PA}) it is straightforward to obtain a useful relation
\begin{align}
	\label{relation}
	2k_0\rho^2={\rm Im}\int_{-\infty}^{\infty}U_\rho(x)|\phi_\rho|^2dx.
\end{align}
It provides   a mathematical expression of the obvious fact that perfectly absorbed or emitted currents cannot be obtained in conservative systems [Im$\,U_\rho(x)\neq 0$ is a necessary conditions for existence of such solutions]. More generally, it follows from (\ref{relation}) that $\PT$-symmetric potentials, for which $U_\rho(x)=\PT U_\rho(x)=U_\rho^*(-x)$, do not support $\PT$-symmetric  currents characterized by  $|\phi_\rho(x)|^2=|\phi_\rho(-x)|^2$.  \rerev{From the physical point of view, Eq.~(\ref{relation}) can be considered as a balance equation which relates the energy dissipated by the localized absorbing potential (or emitted by the localized gain) and the strength of the energy fluxes incoming from the infinity (or outgoing towards the infinity).}

\section{Continuation   preserving the   gain-and-loss distribution}
\label{sec:Wadati}

It is known~\cite{ZezKon20} that under certain (and not very restrictive) conditions, a linear SS-solution $\phi_\s(x)$ exists if and only if the potential $U_0(x)$ in Eq.~(\ref{eq:linear}) admits the representation
\begin{equation}
	\label{eq:wadati}
	\rev{	U_0(x) = -w^2(x) - i\frac{dw(x)}{dx} + k_0^2},
\end{equation}
where $w(x)=w_{re}(x) + iw_{im}(x)$, with $w_{re}(x)=$Re$\,w(x)$ and  $w_{im}(x)=$Im$\,w(x)$, is a complex-valued base function with    asymptotic behavior
\begin{align}
	\label{asymp}
	\lim_{x\to \pm \infty } w(x) = \mp k_0.
\end{align}
If additionally $w(x)$ is a continuous function of $x$, then the linear  solution allows for a universal representation through the base function 
\begin{equation}
	\label{eq:uni}
	\phi_\s(x)=   c_\s \exp\left[-i\int_{x_\s}^xw(\xi) d\xi\right],
\end{equation}
where $x_\s$ is an arbitrary point of the real axis, and constant $c_\s$  is introduced to ensure the boundary conditions (\ref{bound-lin-eq}):
\begin{equation}
\label{eq:c0}
	c_\s  = \exp\left[\int_{-\infty}^{x_\s} w_{im}(\xi) d\xi\right]. 
\end{equation}
With some technical amendments, the consideration  can be  generalized on the case when $w(x)$ has discontinuities  of certain form \cite{ZezKon20}. 
 
The universal form (\ref{eq:uni}) can be used for construction of nonlinear currents bifurcating from the linear limit.   Indeed, first we notice that (\ref{bound-lin-eq}) imposes the following constraint on the base function
\begin{equation}
	\label{eq:cond}
	\int_{-\infty}^{\infty} w_{im}(x) dx = 0.
\end{equation}
Next, we observe that solution $\phi_\rho(x) = \rho\phi_\s(x)$, where  $\phi_\s(x)$ is defined by (\ref{eq:uni}), satisfies the nonlinear    equation (\ref{eq:nonlin}) with the  deformed potential 
\begin{equation}
\label{eq:deform}
U_\rho(x)=U_0(x) +  \rho^2\left(1-\exp\left[-2\int_{x}^\infty w_{im}(\xi)d\xi \right] \right).
\end{equation}

Thus, constraint (\ref{eq:cond}) can be viewed as a {\em sufficient condition} for a possibility of continuation of a linear SS-solution without zeros on the real axis [i.e., the one allowing the representation (\ref{eq:uni})] into the nonlinear domain by a continuous deformation of the potential (\ref{eq:deform}), where $\rho$ is a bifurcation parameter. Condition  (\ref{eq:cond}) is rather general: say, it holds if the potential $U_0(x)$ is an even function.   Deformation (\ref{eq:deform})  only affects the real part of the potential, while the imaginary part of $U_\rho(x)$    does not depend on $\rho$.   In a special case when $w(x)$ is purely real,  i.e., $w_{im}(x)\equiv 0$, no deformation of the potential is necessary at all.  This reflects the fact that the respective solution
$\phi_\rho(x)=\rho\phi_\s(x)$ has a constant amplitude, i.e., $|\phi(x)| \equiv \rho$. 

As we mentioned above, if
\begin{align}
	 \label{eq:cond_1}
	 \int_{-\infty}^{\infty} w_{im}(x) dx =\gamma \neq 0,
\end{align}
then no weak deformation of the respective linear potential   $U_0(x)$ admits continuation of the linear SS-solution to nonlinear regime. In the meantime, allowing spatially unconfined deformations one can define a meaningful small-amplitude limit of the respective       nonlinear currents. Indeed, 
let $f(x)$ be an arbitrary sufficiently well-behaved real-valued function such that 
 \begin{align}
	\label{distr}
	\int_{-\infty}^{\infty}f(x)dx=\gamma. 
\end{align}
Define 
\begin{align}
	\hw(x)=w(x) - i\rho^2f(\rho^2 x). 
\end{align}
Obviously $\hw(x)$ satisfies (\ref{eq:cond}), i.e.,
\begin{align}
	{\rm Im}\int_{-\infty}^\infty\hw(x)dx=0
\end{align}
and has the required asymptotic behavior (\ref{asymp}). We therefore can use this function  to construct a nonlinear SS-solution using expressions (\ref{eq:uni})--(\ref{eq:c0}), where $w(x)$ is to be replaced with $\hw(x)$. 
Respectively, the deformation of the   potential  is given as 
\begin{align}
	U_\rho(x)=&-\hw^2- i\hw_x + k_0^2 
	\nonumber \\
	&+ \rho^2\left(1-\exp\left[-2{\rm Im}\int_{x}^\infty \hw(\xi)d\xi \right] \right)
\end{align}	 
Notice however that the deformation of the potential is nonlocal in the sense that the region of its localization, that can be estimated as $|x|\lesssim1/\rho$, tends to infinity as $\rho$ approaches zero.

\section{Continuation   preserving   the shape of the potential: case study for the rectangular barrier}
\label{sec:rectang}

We have demonstrated in the previous section that any linear SS-solution with equal background amplitudes  at the infinities can be continued to  the nonlinear regime with a proper deformation of only real part of the linear potential [see (\ref{eq:deform})]. Since deformations of the imaginary part of a linear potential can also be considered, now, using a particular example of a rectangular potential, we  demonstrate that continuation of the SS-solutions to the nonlinear domain is not unique.  To this end, we define the rectangular function 
\begin{equation}
\label{eq:rect}
\rect(x) :=   \left\{ 
\begin{array}{ll}
1  &\mbox{if } |x| \leq 1
\\[2mm]
0 & \mbox{if }  |x| \geq 1,
\end{array}
\right.      
\end{equation}
and consider the potential 
\begin{equation}
\label{eq:U}
U_\rho(x)=  \zeta_\rho\, \rect(x),    
\end{equation}
where  $\zeta_\rho$ is a complex  coefficient. As before,   we use   subscript  $\rho$ to parameterize the deformation of the potential for different values of the background amplitude, and  $\rho=0$ corresponds to the  linear limit. Hence the potential (\ref{eq:U}) remains rectangular upon the change of   $\zeta_\rho$.  

\subsection{Linear spectral singularities}

The linear SS-solutions of (\ref{eq:linear}) with the rectangular  potential  have  already been addressed in literature (see e.g.  \cite{CPA1,CPA2,CPA4,ZezKon20,Most_rect09,Most_rect11,Most_rect15,Most_rect17}). For  completeness, herein we   recall some of the known results. Consider the linear problem (\ref{eq:linear}) with rectangular potential  $U_0(x)$  defined by (\ref{eq:U}) with $\zeta_\rho=\zeta_0$. In order to identify the corresponding SS-solutions, it is convenient to use the representation $\zeta_\s = k_0^2 - \kappa_\s^2$, where $\kappa_\s$ is a new complex parameter. Then,  under the condition 
\begin{equation}
\label{eq:k0even}
k_0=i\kappa_\s\tan\kappa_\s,
\end{equation}
there exists an even SS-solution 
\begin{equation}
\label{eq:even}
\phi_\s^{even}(x)=  \left\{ 
\begin{array}{ll}
e^{ik_0(|x|-1)} & \mbox{if } |x| \geq 1
\\[2mm]
\cos(\kappa_\s x)/\cos \kappa_\s  & \mbox{if } |x| \leq 1.
\end{array}
\right.      
\end{equation}
\rev{This SS-solution and the corresponding potential  $U_0(x)$ can be represented in the form (\ref{eq:uni})  and (\ref{eq:wadati}) with the base function defined as \cite{ZezKon20}
\begin{equation}
w(x)=  \left\{ 
\begin{array}{ll}
\mp k_0 & \mbox{if } \pm x  \geq 1
\\[2mm]
-i\kappa_\s\tan(\kappa_0 x) & \mbox{if } |x| \leq 1.
\end{array}
\right.      
\end{equation}}

On the other hand,  under the condition  
\begin{equation}
\label{eq:k0odd}
k_0=-i\kappa_\s\cot\kappa_\s
\end{equation}
 there is  an odd SS-solution
\begin{equation}
\label{eq:odd}
\phi_\s^{odd}(x)=  \left\{ 
\begin{array}{ll}
{\pm}   e^{ ik_0(|x|-1)} & \mbox{if } \pm x > 1
\\[2mm]
\sin(\kappa_\s x)/\sin \kappa_\s  &\mbox{if } |x| \leq 1.
\end{array}
\right.  
\end{equation}
\rev{For the odd solution the base function in  (\ref{eq:uni})  and (\ref{eq:wadati}) has the form
	\begin{equation}
	w(x)=  \left\{ 
	\begin{array}{ll}
	\mp k_0 & \mbox{if } \pm  x \geq 1
	\\[2mm]
	 i\kappa_\s\cot(\kappa_0 x) & \mbox{if } |x| \leq 1.
	\end{array}
	\right.      
	\end{equation}}

\begin{figure}
	\begin{centering}%
		\includegraphics[width=0.99\columnwidth]{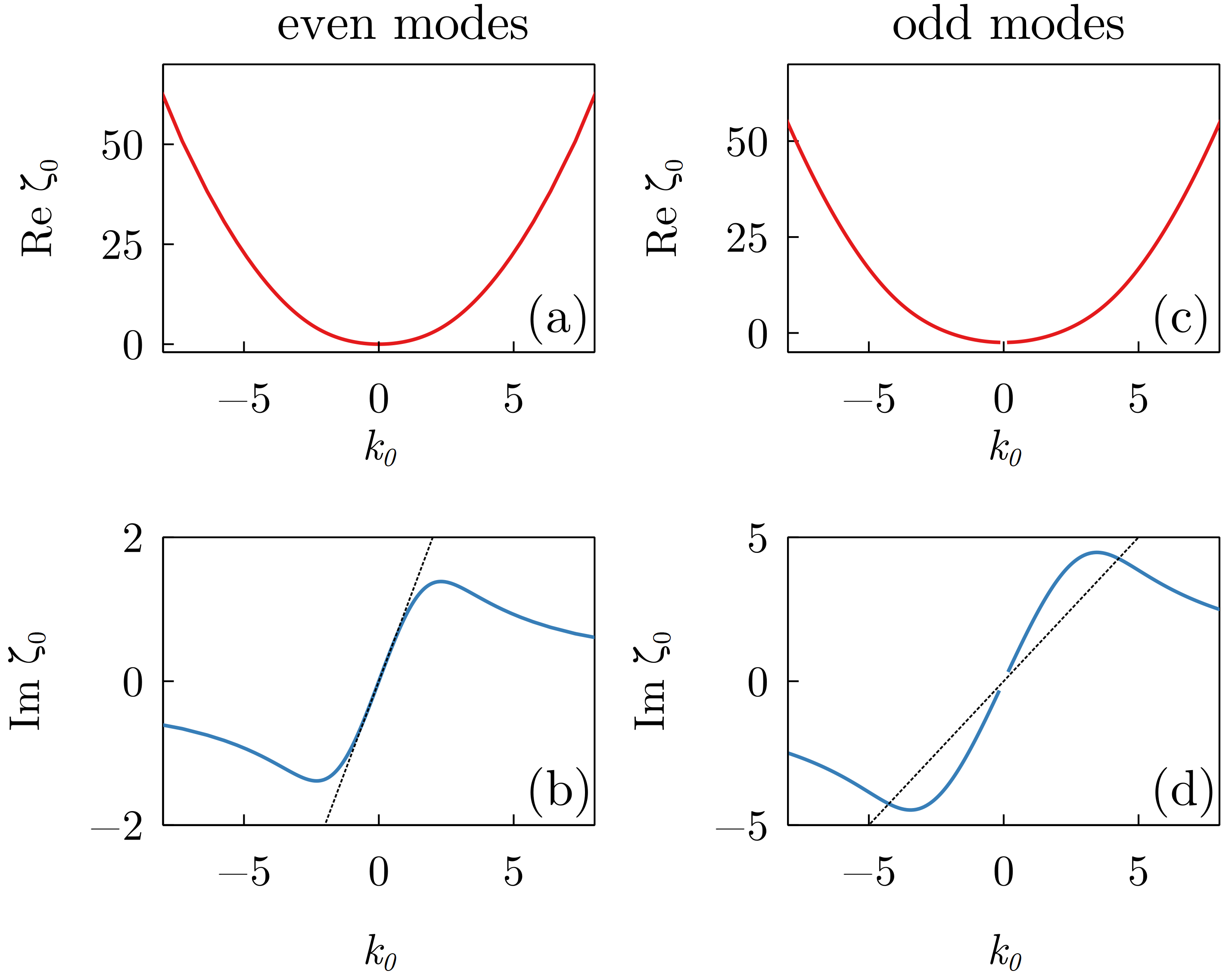}		\par
	\end{centering}
	\caption{\rev{Complex coefficient} $\zeta_0$ of rectangular potential  corresponding to spectral singularities in the linear limit for   even (a,b) and odd (c,d) modes. 
		 Positive and negative   $k_0$ correspond to lasing and absorption, respectively. Dotted lines in (b,d) correspond  to $\IM\zeta_\s=k_0$.  
	  }
	\label{fig:zero}
\end{figure}

Solving   Eqs.~(\ref{eq:k0even}) and (\ref{eq:k0odd}), we find that for each $k_0\ne 0$ there exist one even and one odd linear mode. Dependencies of the complex \rev{parameter} $\zeta_0$  on $k_0$ for linear spectral singularities are shown in Fig.~\ref{fig:zero}. We notice that for even modes the inequality $|\IM\zeta_\s| < |k_0|$ is valid. For odd modes the latter inequality holds only for wavenumbers of sufficiently large \rev{modulus}, whereas for small $k_0$ the opposite inequality takes place, i.e.,   $|\IM\zeta_\s| > |k_0|$.

\subsection{Bifurcations from  linear SS-solutions}

Let $\phi_\s(x)$ be the linear SS-solution  defined either by Eq.~(\ref{eq:even}) or by Eq.~(\ref{eq:odd}).  Apart from $\phi_0$, linear equation (\ref{eq:linear}) has  a linearly independent  solution $\tphi_\s(x)$ which can be written as
\begin{align}
	\tphi_\s(x) = \phi_\s(x)\int_0^x  \phi_\s^{-2}(\xi)\, d\xi.
\end{align}
The solutions $\phi_\s(x)$ and $\tphi_\s(x)$ are of opposite parities and their Wronskian is identically equal to unity: $\phi_\s \tphi_{\s,x} - \phi_{\s,x}\tphi_\s \equiv1$. 

In order   to construct a continuation of the SS-solution  $\phi_\s$ for the nonlinear equation (\ref{eq:nonlin}), we treat $\rho$ as a small parameter  and introduce  the following asymptotic expansions for the solution and for the \rev{coefficient} of the rectangular potential:
\begin{subequations}
	\label{eq:expansions}
\begin{eqnarray}
\phi_\rho(x) = \rho\phi_\s(x) + \sum_{m=1}^\infty \rho^{2m+1}q_m(x),\\
\zeta_\rho=\zeta_0  +\sum_{m=1}^\infty \rho^{2m} z_m,
\end{eqnarray}
\end{subequations}
where $q_m(x)$ and $z_m$ are  coefficients to be found. Formal existence of   asymptotic expansions (\ref{eq:expansions}) for all orders of $m$ can be established in the straightforward way. Indeed,  substituting  (\ref{eq:expansions}) into the nonlinear equation (\ref{eq:nonlin}) and collecting the terms with equal powers of $\rho$, we observe that in  $\rho$- and $\rho^2$-orders   the resulting equations are satisfied automatically. The first nontrivial equation arises   at $\rho^3$-order   and reads
\begin{align}
\label{eq:3}
[\partial_x^2 + k_0^2 - U_\s(x)]q_1(x) + f_1(x)=0,
\end{align}
where
\begin{align}
\label{eq:31}
f_1(x) = [1 -  |\phi_0(x)|^2  - z_1\, \rect(x)]\phi_0(x).
\end{align}
A general solution of Eq.~(\ref{eq:3})   can be written in the form 
\begin{eqnarray*}
q_1(x) =  C_1\phi_\s(x) + C_2 \tphi_\s(x)  -  \tphi_\s(x) \int_0^x\phi_\s(\xi)f_1(\xi)d\xi \nonumber \\
+\phi_\s(x)\int_0^x \tphi_\s(\xi)f_1(\xi) d\xi,
\end{eqnarray*}
where $C_{1,2}$ are, so far, arbitrary constants. Looking for a solution that has the same parity  as   $\phi_\s(x)$ does, we set $C_2=0$.  Since the function  $f_1(x)$ is identically zero outside the interval $[-1, 1]$, we can   define the integrals  
\begin{eqnarray}
\lambda_1 =  \int_0^\infty\phi_\s(\xi)f_1(\xi)d\xi = \int_0^1\phi_\s(\xi)f_1(\xi)d\xi,\\
\tl_1 = \int_0^\infty\tphi_\s(\xi)f_1(\xi) dx =\int_0^1\tphi_\s(\xi)f_1(\xi) dx .
\end{eqnarray}
Looking for a localized function $q_1(x)$,  we require  $\lambda_1=0$ and set $C_1=-\tl_1$. The requirement   $\lambda_1=0$     determines the first correction to the rectangular potential:
\begin{equation}
\label{eq:zeta1}
z_1 = \frac{\int_0^1  (1 - |\phi_\s|^2)\phi_\s^2d x}{\int_0^1\phi_\s^2 d x}.
\end{equation} 
This quotient is generically complex, i.e., both real and imaginary part of the rectangular  potential should be adjusted.

A similar procedure can be carried out for all higher powers of $\rho$ which determine further corrections to the complex  \rev{coefficient} of the potential.   The correction $q_1(x)$ constructed in this way, as well as all higher-order corrections  $q_m$ are identically zero outside the interval $[-1, 1]$.  Therefore a nonlinear perfectly absorbing or lasing solution constructed through this procedure coincide with the properly scaled linear SS-solution outside the support of the potential, i.e., 
{$\phi_\rho(x)-\rho \phi_0(x)\equiv 0$}
as $|x|\geq 1$. This in particular means that $|\phi_\rho(x)|\equiv \rho$ as $|x|\geq 1$.

\subsection{Numerical study of nonlinear currents} 

\begin{figure}
	\begin{centering}%
		\includegraphics[width=0.99\columnwidth]{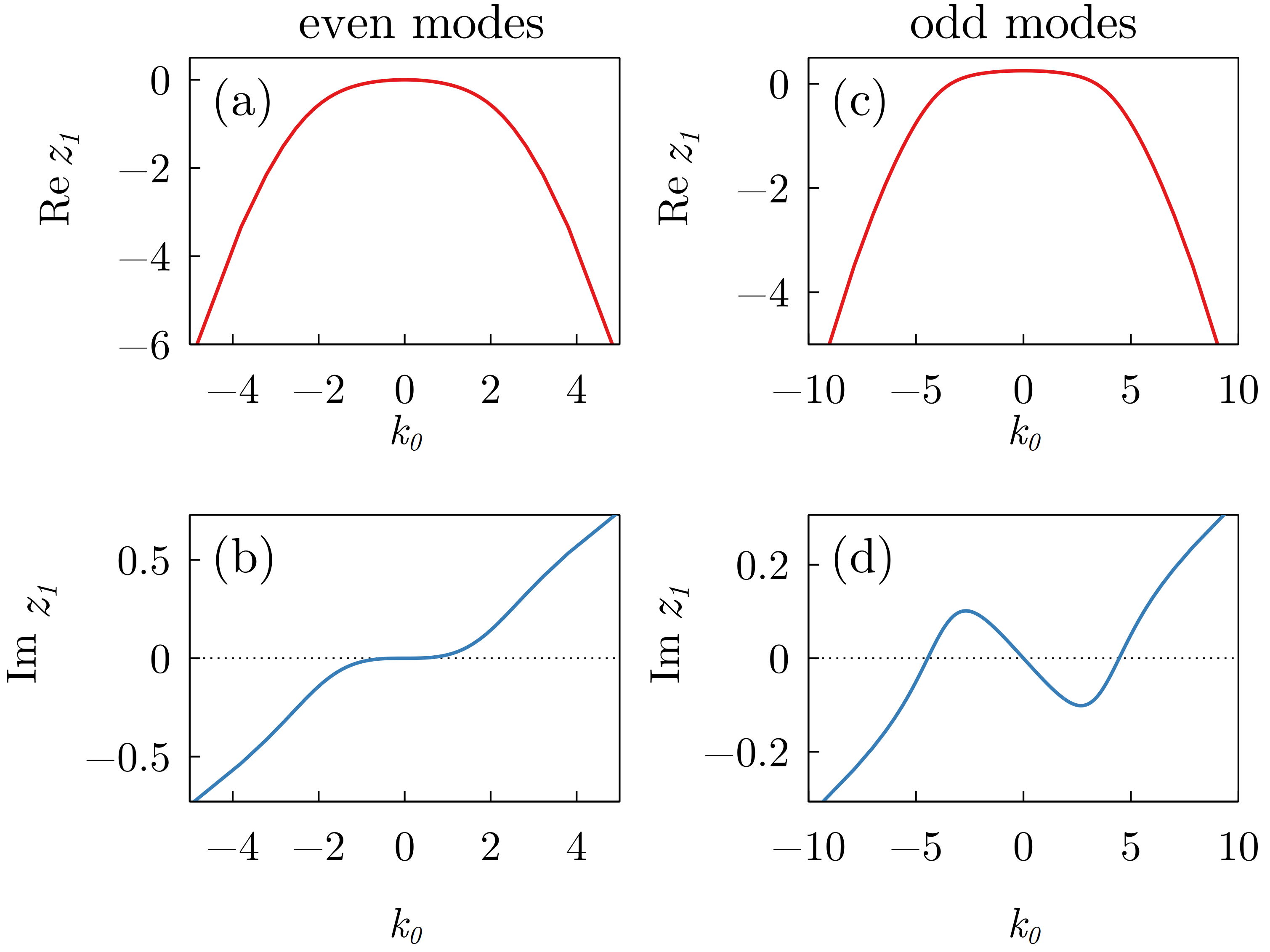}		\par
	\end{centering}
	\caption{Real and imaginary parts of the leading correction $z_1$ to the complex  \rev{coefficient} of the rectangular potential that features a SS in the linear limit \textit{vs.}    $k_0$. 	}
	\label{fig:one}
\end{figure}

Using  (\ref{eq:zeta1}), we computed the leading correction $z_1$ to the  \rev{coefficient} of the rectangular potential. The result is presented in Fig.~\ref{fig:one}   as a  function of $k_0$. For even modes, illustrated in Fig.~\ref{fig:one}(a,b), the imaginary part of $z_1$ has the same sign as  $k_0$. Since positive and negative imaginary part of the complex  \rev{coefficient} $\zeta_\rho$ correspond, respectively, to gain and loss, this means that lasing of weakly nonlinear even modes requires stronger gain, than its purely linear counterpart does, while absorption of nonlinear even modes  requires stronger losses than those required by the linear CPA.  

For odd modes illustrated in Fig.~\ref{fig:one}(c,d), the behavior of the leading correction  $z_1$ is different. For small wavenumbers $k_0$ the imaginary part of the coefficient $z_1$  is \emph{negative} for emitting  modes and is \emph{positive} for absorbing modes. In this interval of   wavenumbers, the perfect absorption and lasing of nonlinear modes require, respectively, weaker losses and gain than those of the linear SS-solutions. However, for larger $|k_0|$  the sign of the imaginary part of $z_1$  coincides with the sign of $k_0$.

\begin{figure}
	\begin{centering}%
		\includegraphics[width=0.99\columnwidth]{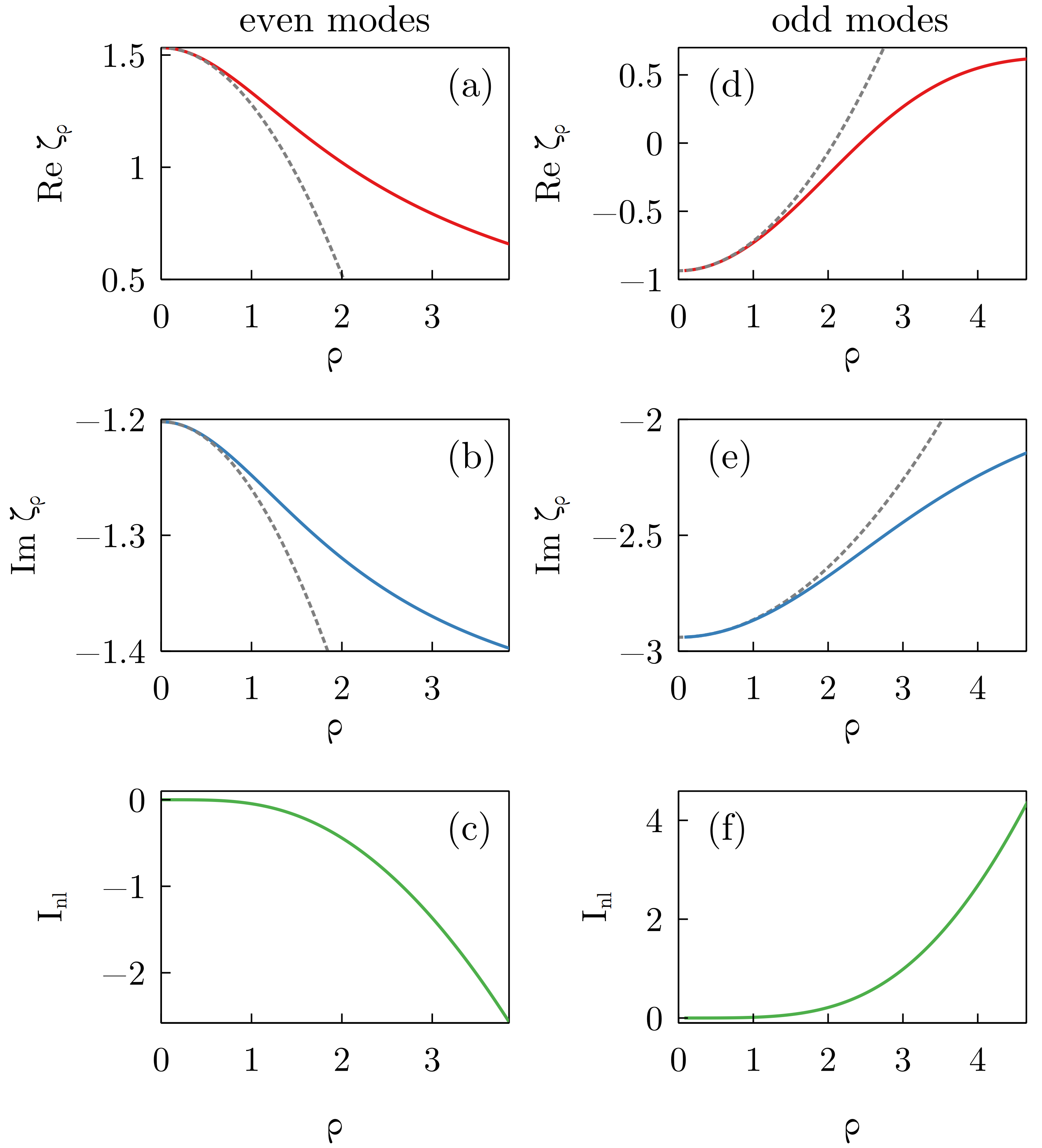}		\par
	\end{centering}
	\caption{Branches of nonlinear perfectly absorbing currents bifurcating from the linear limit which corresponds to zero background amplitude $\rho=0$. Upper and middle rows show real and imaginary parts of the complex  \rev{coefficient} $\zeta_\rho$ of the rectangular potential. Grey dashed lines plot asymptotic predictions $\zeta_\rho  = \zeta_0 + z_1  \rho^2$ 	where $z_1$ is computed from (\ref{eq:zeta1}). Lower panels (c) and (f) show the dependence of $I_{nl}$ on $\rho$.  Even and odd modes correspond, respectively,  to $k_0\approx -1.474$ and  $k_0\approx -1.589$. }
	\label{fig:two}
\end{figure}

\begin{figure}
	\begin{centering}%
		\includegraphics[width=0.99\columnwidth]{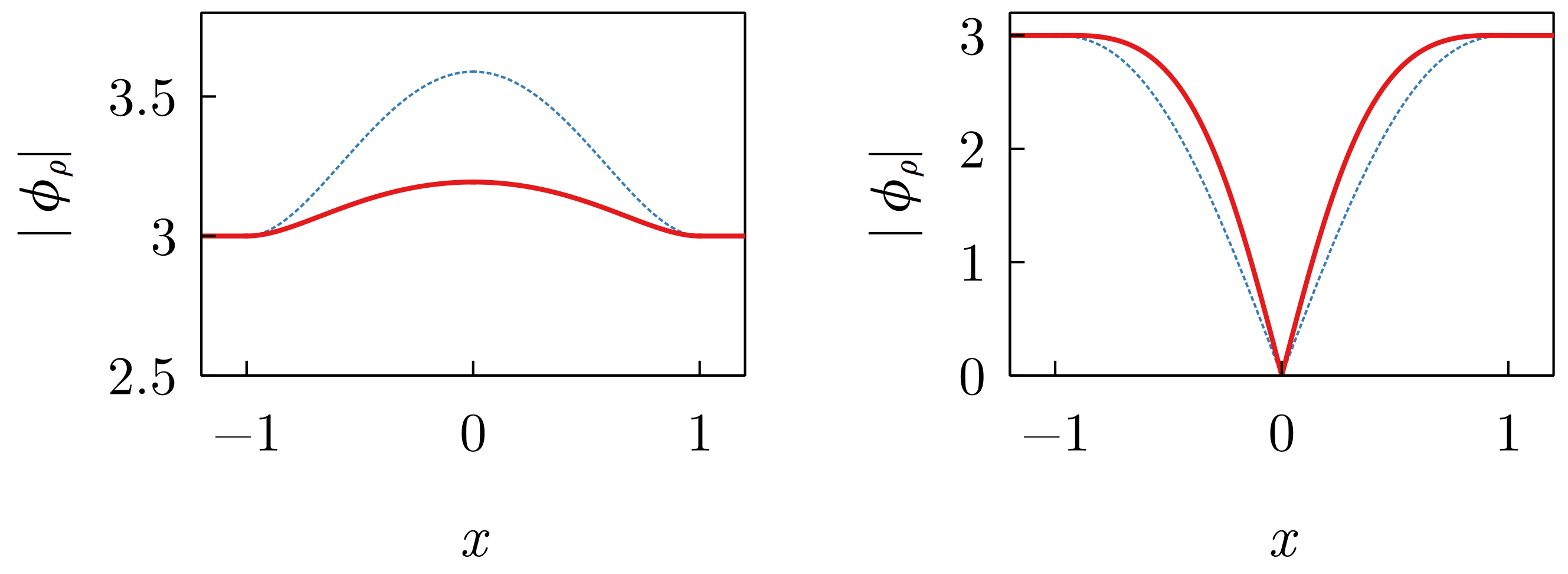}		\par
	\end{centering}
	\caption{Moduli $|\phi_\rho(x)|$ of even and odd nonlinear perfectly absorbed currents   (red curves)   selected from Fig.~\ref{fig:two}  at $\rho=3$ juxtaposed to moduli of their linear counterparts $\rho|\phi_0(x)|$ which are normalized to have the same  background amplitude (blue dashed lines). }
	\label{fig:three}
\end{figure}

To validate the predictions  of the asymptotic expansions (\ref{eq:expansions}), we have computed numerically several branches of perfectly absorbed nonlinear modes bifurcating from the linear limit. Typical results for even and odd  currents are shown in Fig.~\ref{fig:two}(a-c) and Fig.~\ref{fig:two}(d-f), respectively. In the vicinity of the bifurcations from the linear limit at $\rho=0$, the real and imaginary parts of the complex  \rev{coefficient} $\zeta_\rho$ are well approximated by the analytical prediction (dashed lines). To describe the nonlinearity-induced deviation between the shapes of linear and nonlinear currents, for each amplitude $\rho$    we compute the integral measure
\begin{equation}
I_{nl} = \int_0^1 (|\phi_\rho|^2 - \rho^2 |\phi_0|^2)dx.
\end{equation}
This quantity, plotted in the lower panels of Fig.~\ref{fig:two}, is negative for even modes and positive for odd modes.   Figure~\ref{fig:three} compares the shapes of nonlinear SS solutions with their linear counterparts.

While in the linear context CPA and lasing can be considered as equivalent modulo to time-reversal \cite{Stone}, the corresponding nonlinear regimes are not equivalent from the point of view of stability. \rerev{Here we examine the  linear  stability of perfectly absorbed absorbed currents using the standard analysis \cite{Yang}. A numerical study of linear stability spectra  indicates that   even modes from Fig.~\ref{fig:two} are stable both in the lasing and absorbing cases, whereas odd modes are stable only in the absorbing regime.}

\section{Essentially nonlinear perfectly absorbed currents}
\label{sec:nonlin-currents}

A characteristic property of   currents constructed in the previous section is that the field amplitude of solutions (both linear and nonlinear) is exactly constant outside the support of the rectangular potential, i.e., $|\phi_\rho(x)|\equiv \rho$ if $|x|\geq 1$. Apart from these solutions,   the nonlinear medium supports a different type of perfectly absorbed currents  for which the background density is approached asymptotically, i.e., $\lim_{x\to\pm\infty} |\phi_\rho(x)|=\rho$. While currents of this type have been introduced in \cite{ZKBO} for a purely imaginary localized potential, their eventual relation to linear SSs and the behavior near to the linear limit have not been addressed yet. To clarify these issues,  here   we numerically investigate this type of nonlinear perfectly absorbed currents considering the rectangular potential (\ref{eq:U}).  We will demonstrate that the perfectly absorbed solutions with the asymptotically approached background amplitude   cannot be reduced to the linear SS-solutions. Therefore, the nonlinear perfectly absorbed currents in the rectangular potential can be   classified in two different types:  solutions of the first type [described in the previous section~\ref{sec:rectang}] bifurcate from the spectral singularities in the linear limit, while solutions of the second type [considered in the present section] do not admit the linear limit. 

\begin{figure}
	\begin{centering}%
		\includegraphics[width=0.99\columnwidth]{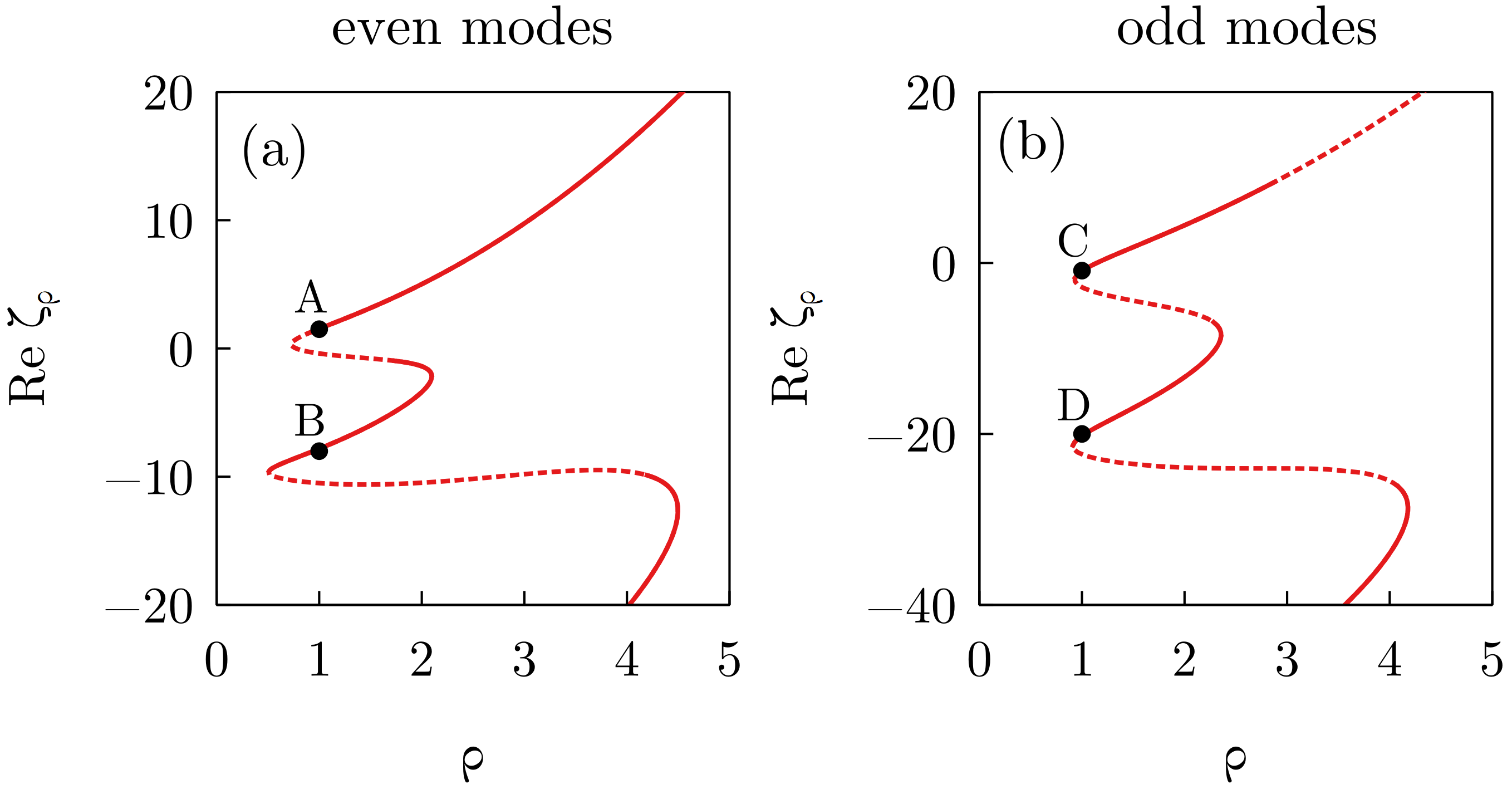}		\par
	\end{centering}
	\caption{Dependencies of $\RE \zeta_\rho$ on $\rho$ for even and odd nonlinear perfectly absorbed currents   without linear counterpart, while $k_0$ and $\IM \zeta_\rho$ remain fixed. For even modes $k_0\approx -0.217$  and $\IM\zeta_\rho \approx -1.202$.  For odd  modes $k_0\approx -0.291$  and $\IM\zeta_\rho \approx -2.940$.  Solid and dotted fragments correspond to stable and unstable  solutions, respectively. \rev{Points labeled A--D correspond to the specific solutions plotted in Fig.~\ref{fig:five}(A-D).}}
	\label{fig:four}
\end{figure}

\begin{figure}
	\begin{centering}%
		\includegraphics[width=0.99\columnwidth]{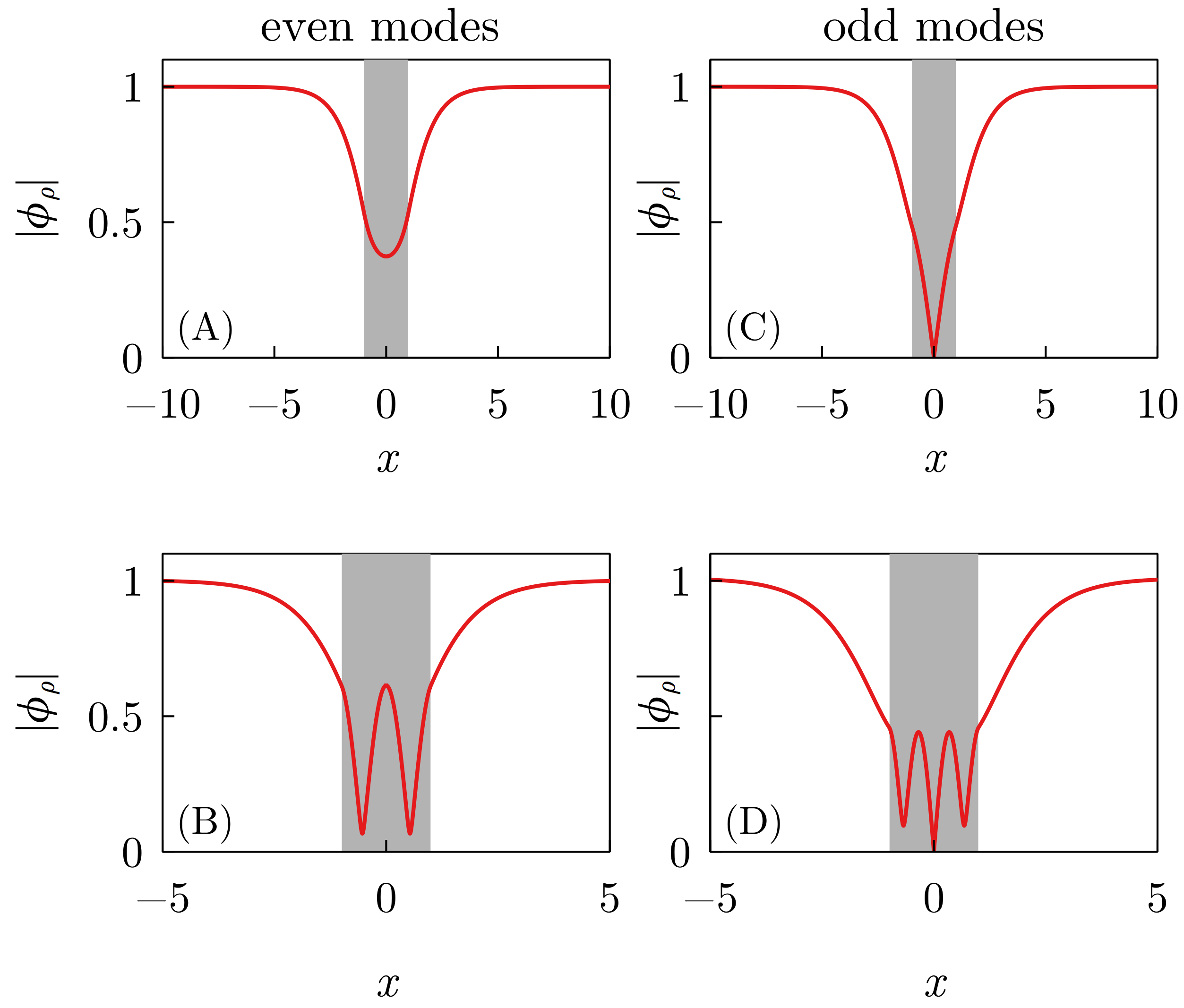}		\par
	\end{centering}
	\caption{Spatial profiles $|\phi_\rho(x)|$ for stable perfectly absorbed currents   with $\rho=1$ from Fig.~\ref{fig:four}. Gray regions correspond  to the domain $x\in [-1, 1]$ there the absorbing potential is confined.  \rev{Panels A--D correspond to the points A--D in Fig.~\ref{fig:four}.} }
	\label{fig:five}
\end{figure}

To compute   nonlinear currents of the second type, we again use the nonlinear equation (\ref{eq:nonlin}) with boundary conditions (\ref{eq:PA}).  In comparison with nonlinear currents  of the first type,  the perfectly absorbed currents of the second type   are easier-to-find, because   perfect absorption at given $k_0$ and background amplitude $\rho$ can be achieved by tuning only  real or  only  imaginary part of the  \rev{coefficient} $\zeta_\rho$. In Fig.~\ref{fig:four} we present diagrams   ($\RE\,\zeta_\rho,\rho$) for the rectangular potentials enabling perfectly absorbed currents at a given $k_0$ and fixed  $\IM\,\zeta_\rho$.  The dependencies feature peculiar snaking behavior, for both even and  odd currents, and, most importantly,  neither dependency can be continued to the linear limit $\rho\to 0$.  Non-existence   of the linear limit for nonlinear currents of this type  agrees with the inequality 
\begin{equation}
\label{eq:bound02}
\rho > \sqrt{2} |k_0|
\end{equation}
which was obtained in \cite{ZKBO} from the analysis of the asymptotic behavior of the spatial profiles of the currents for $x\to \pm \infty$. Spatial profiles of perfectly absorbed states of the second type (exemplified  in Fig.~\ref{fig:five})  are distinctively different from the  currents of   the first type found in Sec.~\ref{sec:rectang}. \rev{ Multiple local minima and maxima of the amplitude for solutions in Fig.~\ref{fig:five}(B,D) can be understood by noticing   that real parts of the corresponding rectangular potentials are negative with large absolute values (see points B and D in Fig.~\ref{fig:four}). In this situation the oscillating shape of the solution inside the rectangular slab can be approximated by the Jacobian elliptic sine: $\phi\approx \sqrt{2m}\nu$sn$(\nu (x-x_0), m)$, where $m$ is the so-called parameter \cite{AS} and the real coefficient $\nu$ characterizes the frequency of spatial oscillations. For a strong rectangular potential one can roughly estimate  $\nu \approx   \sqrt{|\RE \zeta_\rho|/(m+1)}$, which corresponds to fast  spatial oscillations. Qualitatively such a spatial dependence of nonlinear currents resembles that of the linear SS-solutions: see Eqs.~(\ref{eq:even}) and (\ref{eq:odd}), where for $\zeta_0$ with large and negative real part we estimate $\kappa_0 \approx \sqrt{|\RE \zeta_0|}$ (i.e., the same energy in a more deep potential corresponds to smaller wavelength of the eigenmode).}

The linear stability study for nonlinear currents of the second type  reveals several alternating regions with stable (solid lines in Fig.~\ref{fig:four}) and unstable (dashed lines in Fig.~\ref{fig:four}) currents.

\begin{figure}
	\begin{centering}%
		\includegraphics[width=0.99\columnwidth]{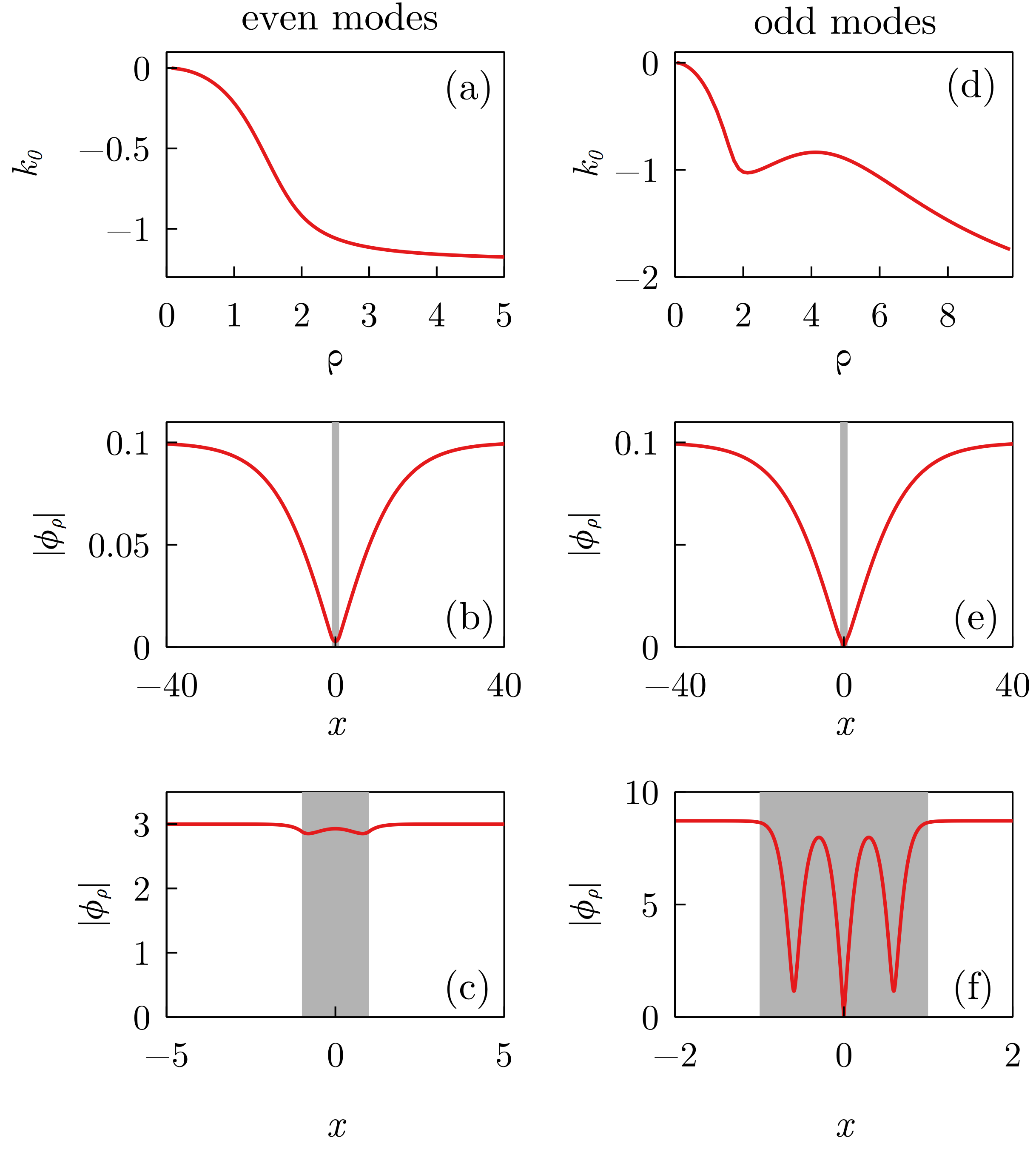}		\par
	\end{centering}
	\caption{(a,d) Dependencies of the wavenumber $k_0$ on the background amplitude $\rho$ for even and odd perfectly absorbed currents of the second type, while the \rev{complex coefficient} $\zeta_\rho$ of the absorbing potential is kept fixed. Panels in two lower rows show spatial profiles $|\phi_\rho(x)|$ for   even and odd modes with different background amplitudes.  Gray regions correspond  to the domain $x\in [-1, 1]$ there the absorbing potential is confined.  Shown solutions are stable. For even and odd currents,   $\zeta\approx 1.533-1.202i$ and $\zeta\approx-0.936-2.940i$, respectively.}
	\label{fig:six}
\end{figure}

So far, our consideration has been focused on deformation of a complex potential for the wavenumber  $k_0$  being fixed both in linear and in nonlinear regimes. Obviously, an alternative statement is also meaningful: one can fix the imaginary potential and consider the  change of $k_0$ under the   variation of the background  amplitude $\rho$. To address this possibility,  we fix real and imaginary parts of the amplitude of the rectangular potential, denoting it by $\zeta$, and consider the  small-amplitude  limit $\rho\to 0$ allowing the wavenumber $k_0$ to change. The corresponding diagrams are presented in Fig.~\ref{fig:six},  where we observe that in the small-amplitude limit  the wavenumber vanishes, $\lim_{\rho\to 0} k_0 =0$, i.e., for small $\rho$ the absorption is effectively inhibited.  Looking to the spatial shapes of the currents with small background amplitude $\rho$,  we notice that the 	attenuation of the absorption is explained by the fact that the   solution  leaves the domain where the absorbing potential is confined [Fig.~\ref{fig:six}, second row]. In the limit of large $\rho$  the amplitude of even states becomes almost constant $|\phi_\rho(x)|\approx \rho$ [Fig.~\ref{fig:six}(c)].  As readily follows from Eq.~(\ref{relation}), for these solutions  $|k_0|$ approaches $|\IM \zeta|$ from below as $\rho\to\infty$.  \rev{The oscillatory behavior of the odd solution in Fig.~\ref{fig:six}(f) can be again explained in terms of   Jacobian functions, because for $\rho\gg 1$ the wavefunction inside the slab can be approximated as  $\phi\approx \sqrt{2m}\nu$sn$(\nu x , m)$, where   $\nu \approx    \sqrt{\rho /(m+1)}$.}

Since for all currents of the second type the inequality $|\phi_\rho(x)|\leq \rho$ holds, using (\ref{relation}) we obtain that  
\begin{align}
\label{eq:bound01}
|k_0| \leq |\IM \zeta_\rho|.
\end{align}
Let us now recall that for linear SSs we have observed that for even modes the inequality $|\IM \zeta_0|<|k_0|$ holds (see Fig.~\ref{fig:zero} and the discussion around).  This leads us to the following conclusion for even currents: if the rectangular potential is chosen such that linear spectral singularity at $k_0$ coexist with nonlinear current of the second type at the same wavenumber $k_0$, then the nonlinear current of the second type requires stronger absorption than the linear current.  Note that  odd currents the situation can be more complex, because in this case the    inequality $|\IM \zeta_0|<|k_0|$ holds only for sufficiently large $k_0$.

\begin{table*}
\begin{tabular}{p{5.5cm}p{5cm}p{4.5cm}}
\textbf{property} &\textbf{currents of the first type} & \textbf{currents of the second type}\\\hline 
linear limit & spectral singularities & does not exist\\[3mm]
field amplitude   
 outside the potential with a finite support & constant  & space-dependent\\[5mm]
existence conditions   at   given wavenumber 
and background amplitude 
& both real 
and imaginary 
parts of the potential must be adjusted  & it is sufficient to adjust only a real or only an imaginary part of the potential\\[3mm]
relation between the wavevector and background amplitude & no \textit{apriori} relation  & $k_0^2< \rho^2/2$
\end{tabular}
\caption{\rerev{Summary of dissimilarities between the two types of nonlinear perfectly absorbed currents supported by the rectangular potential.}}
\label{tbl1}
\end{table*}

\section{Conclusion}
\label{sec:concl}

One-dimensional nonlinear Schr\"odinger equation with a localized complex potential \rev{and spatially-uniform Kerr-type cubic nonlinearity} can support perfectly absorbed or emitted currents. Solutions of this class are characterized by purely incoming or purely outgoing wave boundary conditions and in this respect resemble spectral singularities which are well-known in the linear theory of non-Hermitian operators. Therefore a question emerges on the possibility to transform from a nonlinear current to a spectral singularity (or back) by decreasing (or increasing) the effective nonlinearity strength. This question has been previously answered only for a very specific situation where the amplitude of the solution is constant. In the present  paper, we have   generalized  the earlier treatment to non-constant-amplitude   solutions.  We have demonstrated that in the general situation   bifurcation  of a nonlinear current from a spectral singularity is possible only if the complex potential is deformed properly, and this deformation is not   unique.  For a case example of rectangular potential,   bifurcations from the linear limit are possible only if the complex  \rev{coefficient} of the slab is tuned properly.  In contrast to the   linear case, nonlinear absorbing currents cannot be considered as antilasers, because perfectly absorbing and emitted currents feature different stability properties.  Additionally, the rectangular potential supports essentially nonlinear currents that cannot be reduced to the linear limit. Solutions of this type are very different from those bifurcating from linear spectral singularities. \rerev{The main dissimilarities between the two types of nonlinear currents perfectly absorbed by the rectangular potential are summarized in Table~\ref{tbl1}.}

\begin{acknowledgments}
Work of D.A.Z. is funded by Russian Foundation for Basic Research (RFBR) according to the research
project No. 19-02-00193. V.V.K. acknowledges financial support from the Portuguese Foundation for Science and Technology (FCT) under Contract UIDB/00618/2020.
\end{acknowledgments}

\end{document}